\documentclass[english,aps,manuscript]{revtex4-1}
\usepackage[T1]{fontenc}
\usepackage[latin9]{inputenc}
\usepackage[active]{srcltx}
\usepackage{textcomp}
\usepackage{esint}

\makeatletter
\@ifundefined{textcolor}{}
{%
 \definecolor{BLACK}{gray}{0}
 \definecolor{WHITE}{gray}{1}
 \definecolor{RED}{rgb}{1,0,0}
 \definecolor{GREEN}{rgb}{0,1,0}
 \definecolor{BLUE}{rgb}{0,0,1}
 \definecolor{CYAN}{cmyk}{1,0,0,0}
 \definecolor{MAGENTA}{cmyk}{0,1,0,0}
 \definecolor{YELLOW}{cmyk}{0,0,1,0}
 }

\makeatother

\usepackage{babel}

\begin{document}

\title{Geometrothermodynamics of a Charged Black Hole of String Theory}

\author{Alexis Larrañaga}

\address{National Astronomical Observatory. National University of Colombia.}

\author{Sindy Mojica}

\address{Department of Physics. National University of Colombia.}
\begin{abstract}
The thermodynamics of the Gibbons-Maeda-Garfinkle-Horowitz-Strominger
(GMGHS) charged black hole from string theory is reformulated within
the context of the recently developed formalism of geometrothermodynamics.
The geometry of the space of equilibrium states is curved, but we
show that the thermodynamic curvature does not diverge  when the black hole solution becomes a naked
singularity. This provides a counterexample to the conventional notion
that  a thermodynamical curvature divergence signals the occurrence of a phase
transition. 


\end{abstract}
\maketitle

\section{Introduction}

The thermodynamics of black holes has been studied extensively since
the work of Hawking \cite{key-1}. The notion of critical behaviour
has arisen in several contexts for black holes, ranging from the Hawking-Page
\cite{key-2} phase transition in hot anti-de-Sitter space and the
pioneering work by Davies \cite{key-3} on the thermodynamics of Kerr-Newman
black holes, to the idea that the extremal limit of various black
hole families might themselves be regarded as genuine critical points
\cite{key-4,key-5,key-6}. The use of geometry in statistical mechanics
was pioneered by Ruppeiner \cite{rup79} and Weinhold 
\cite{wei1}, who suggested that the curvature of a metric defined
on the space of parameters of a statistical mechanical theory could
provide information about the phase structure. However, some puzzling
anomalies become apparent when these methods are applied to the study
of black hole thermodynamics. A possible resolution was suggested
by Quevedo\textquoteright{}s geometrothermodynamics (GTD) whose starting
point \cite{quev07} was the observation that standard thermodynamics
was invariant with respect to Legendre transformations, since one
expects consistent results whatever starting potential one takes. 

The formalism of GTD indicates that phase transitions occur at those
points where the thermodynamic curvature is singular. In particular,
these singularities represent critical points where the geometric
description of GTD does not hold anymore, for example, at point where
a naked singularity of spacetime appears. In this paper we apply the
GTD formalism to the Gibbons-Maeda-Garfinkle-Horowitz-Strominger (GMGHS)
charged black hole from string theory to investigate the behaviour
of the thermodynamical curvature.

\section{Geometrothermodynamics in Brief}

The formulation of GTD is based on the use of contact geometry as
a framework for thermodynamics. Consider the $(2n+1)$-dimensional
thermodynamic phase space $\mathcal{T}$ coordinatized by the thermodynamic
potential $\Phi$, extensive variables $E^{a}$, and intensive variables
$I^{a}$, with $a=1,...,n$. Consider on $\mathcal{T}$ a non-degenerate
metric $G=G(Z^{A})$, with $Z^{A}=\{\Phi,E^{a},I^{a}\}$, and the
Gibbs 1-form $\Theta=d\Phi-\delta_{ab}I^{a}dE^{b}$, with $\delta_{ab}={\rm diag}(1,1,...,1)$.
If the condition $\Theta\wedge(d\Theta)^{n}\neq0$ is satisfied, the
set $(\mathcal{T},\Theta,G)$ defines a contact Riemannian manifold.
The Gibbs 1-form is invariant with respect to Legendre transformations,
while the metric $G$ is Legendre invariant if its functional dependence
on $Z^{A}$ does not change under a Legendre transformation. Legendre
invariance guarantees that the geometric properties of $G$ do not
depend on the thermodynamic potential used in its construction. 

The $n$-dimensional subspace $\mathcal{E}\subset\mathcal{T}$ is
called the space of equilibrium thermodynamic states if it is determined
by the smooth mapping \begin{eqnarray}
\varphi:\ \mathcal{E} & \longrightarrow & \mathcal{T}\nonumber \\
(E^{a}) & \longmapsto & \left(\Phi,E^{a},I^{a}\right)
\end{eqnarray}
with $\Phi=\Phi(E^{a})$, and the condition $\varphi^{*}(\Theta)=0$
is satisfied, i.e. \begin{equation}
d\Phi=\delta_{ab}I^{a}dE^{b}\end{equation}
 \begin{equation}
\frac{\partial\Phi}{\partial E^{a}}=\delta_{ab}I^{b}.\end{equation}
The first of these equations corresponds to the first law of thermodynamics,
whereas the second one is usually known as the condition for thermodynamic
equilibrium (the intensive thermodynamic variables are dual to the
extensive ones). Note that the mapping $\varphi$ defined above implies that
the equation $\Phi=\Phi(E^{a})$ must be explicitly given, becoming the fundamental equation from which all the equations
of state can be derived. Finally, the second law of thermodynamics
is equivalent to the convexity condition on the thermodynamic potential,
\begin{equation}
\partial^{2}\Phi/\partial E^{a}\partial E^{b}\geq0.\end{equation}

The thermodynamic potential satisfies the homogeneity condition $\Phi(\lambda E^{a})=\lambda^{\beta}\Phi(E^{a})$
for constant parameters $\lambda$ and $\beta$. Therefore, it satisfies
the Euler's identity, \begin{equation}
\beta\Phi(E^{a})=\delta_{ab}I^{b}E^{a},\end{equation}
and using the first law of thermodynamics, we obtain the Gibbs-Duhem
relation, \begin{equation}
(1-\beta)\delta_{ab}I^{a}dE^{b}+\delta_{ab}E^{a}dI^{b}=0.\end{equation}

We also define a non-degenerate metric structure $g$ on ${\cal E}$,
that is compatible with the metric $G$ on ${\cal T}$. We will use
the pullback $\varphi^{*}$ to define $g$ such that it is induced
by $G$ as $g=\varphi^{*}(G)$. As is shown in \cite{quev07}, a Legendre
invariant metric $G$ induces a Legendre invariant metric $g$. There
is a vast number of metrics on ${\cal T}$ that satisfy the Legendre
invariance condition. For instance, the metric structure \begin{equation}
G=\Theta^{2}+(\delta_{ab}E^{a}I^{b})(\delta_{cd}dE^{c}dI^{d})\label{eq:Gmetric}\end{equation}
 is Legendre invariant (because of the invariance of the Gibbs 1-form)
and induces on ${\cal E}$ the Quevedo's metric \begin{equation}
g=\Phi\left(E^{c}\right)\frac{\partial^{2}\Phi}{\partial E^{a}\partial E^{b}}dE^{a}dE^{b}.\label{eq:metricg}\end{equation}

The geometry described by the metric $g=\varphi^{*}(G)$ is invariant
with respect to arbitrary diffeomorphisms performed on ${\cal E}$.

Now, Weinhold's metric $g^{W}$ is defined as the Hessian in the mass
representation \cite{wei1}, whereas Ruppeiner's metric $g^{R}$ is
given as minus the Hessian in the entropy representation \cite{rup79},
\begin{eqnarray}
g^{W} & = & \frac{\partial^{2}M}{\partial E^{a}\partial E^{b}}dE^{a}dE^{b}\\
g^{R} & = & -\frac{\partial^{2}S}{\partial F^{a}\partial F^{b}}dF^{a}dF^{b},\label{eq:rupp}
\end{eqnarray}

where $E^{a}=\{S,Q\}$ and $F^{a}=\{M,Q\}$. As is well known, \cite{quev07},
Weinhold's and Ruppeiner's metrics are not Legendre invariant and
from the analysis given above, it is clear that these metrics must
be related by $g^{W}=\left(\frac{\partial S}{\partial M}\right)^{-1}g^{R}=Tg^{R}$.

In GTD the simplest way to reach the Legendre invariance for $g^{W}$
is to apply a conformal transformation with the thermodynamic potential
as the conformal factor, as given in equation (\ref{eq:metricg}),
\begin{equation}
g=M\frac{\partial^{2}M}{\partial E^{a}\partial E^{b}}dE^{a}dE^{b}=Mg^{W},\label{eq:invariantg2}\end{equation}
 or, using equation (\ref{eq:rupp}), it can be written in terms of
the Ruppeiner's metric as \begin{equation}
g=-M\left(\frac{\partial S}{\partial M}\right)^{-1}\frac{\partial^{2}S}{\partial F^{a}\partial F^{b}}dF^{a}dF^{b}=MTg^{R}.\label{eq:invariantg}\end{equation}

\section{The GMGHS Black Hole}

The low energy effective action of the heterotic string theory in
four dimensions is given by
\begin{equation}
\mathcal{A}=\int d^{4}x\sqrt{-\tilde{g}}e^{-\psi}\left(-R+\frac{1}{12}H_{\mu\nu\rho}H^{\mu\nu\rho}-\tilde{G}^{\mu\nu}\partial_{\mu}\psi\partial_{\nu}\psi+\frac{1}{8}F_{\mu\nu}F^{\mu\nu}\right),\end{equation}

where $R$ is the Ricci scalar, $\tilde{G}_{\mu\nu}$ is the metric
that arises naturally in the $\sigma$ model, 

\begin{equation}
F_{\mu\nu}=\partial_{\mu}A_{\nu}-\partial_{\nu}A_{\mu}\end{equation}
is the Maxwell field associated with a $U\left(1\right)$ subgroup
of $E_{8}\times E_{8}$, $\psi$ is the dilaton field and

\begin{equation}
H_{\mu\nu\rho}=\partial_{\mu}B_{\nu\rho}+\partial_{\nu}B_{\rho\mu}+\partial_{\rho}B_{\mu\nu}-\left[\Omega_{3}\left(A\right)\right]_{\mu\nu\rho},\end{equation}
where $B_{\mu\nu}$ is the antisymmetric tensor gauge field and

\begin{equation}
\left[\Omega_{3}\left(A\right)\right]_{\mu\nu\rho}=\frac{1}{4}\left(A_{\mu}F_{\nu\rho}+A_{\nu}F_{\rho\mu}+A_{\rho}F_{\mu\nu}\right)\end{equation}
is the gauge Chern-Simons term. Considering $H_{\mu\nu\rho}=0$ and
working in the conformal Einstein frame, the action becomes

\begin{equation}
\mathcal{A}=\int d^{4}x\sqrt{-\tilde{g}}\left(-R+2\left(\nabla\psi\right)^{2}+e^{-2\phi}F^{2}\right),\end{equation}
where the Einstein frame metric $\tilde{g}_{\mu\nu}$ is related to
$\tilde{G}_{\mu\nu}$ through the dilaton,

\begin{equation}
\tilde{g}_{\mu\nu}=e^{-\psi}\tilde{G}_{\mu\nu}.\end{equation}

The charged black hole solution, known as the Gibbons-Maeda-Garfinkle-Horowitz-Strominger
(GMGHS) solution, is given by \cite{gmghs,shaoWEn}

\begin{eqnarray}
ds^{2} & = & -\left(1-\frac{2M}{r}\right)dt^{2}+\left(1-\frac{2M}{r}\right)^{-1}dr^{2}\nonumber \\
 &  & +r\left(r-\frac{Q^{2}e^{-2\psi_{0}}}{M}\right)d\Omega^{2}\\
e^{-2\psi} & = & e^{-2\psi_{0}}\left(1-\frac{Q^{2}}{Mr}\right)\\
F & = & Q\sin\theta d\theta\wedge d\vartheta
\end{eqnarray}
where $M$ is the mass of the black hole, $Q$ the electric charge
and $\psi_{0}$ is the asymptotic value of the dilaton. In addition
to their mass $M$ and the charge $Q$, the GMGHS solution is also
characterized by the dilaton charge

\begin{equation}
D=-\frac{Q^{2}e^{-2\varphi_{0}}}{2M}.\end{equation}

Note that this metric become Schwarzschild\textasciiacute{}s solution
if $Q=0$ and has a spherical event horizon at

\begin{equation}
r_{H}=2M\label{eq:horizon}\end{equation}

with an area given by

\begin{eqnarray}
A & = & 4\pi r_{H}\left(r_{H}-\frac{Q^{2}e^{-2\psi_{0}}}{M}\right)\\
A & = & 4\pi r_{H}^{2}-8\pi Q^{2}e^{-2\psi_{0}}.\label{eq:area}
\end{eqnarray}

Equation (\ref{eq:area}) tell us that the area of the horizon goes
to zero if

\begin{equation}
r_{H}^{2}=2Q^{2}e^{-2\psi_{0}},\end{equation}
i.e. the GMGHS solution becomes a naked singularity when

\begin{equation}
M^{2}\leq\frac{1}{2}Q^{2}e^{-2\psi_{0}}.\end{equation}

The Hawking temperature is

\begin{equation}
T=\frac{\kappa}{2\pi}=\frac{1}{8\pi M}\end{equation}

which is independent of charge. Finally, the electric potential computed
on the horizon of the black hole is 

\begin{equation}
\phi=\frac{Q}{r_{H}}e^{-2\psi_{0}},\end{equation}

while the entropy is

\begin{equation}
S=\frac{A}{4}=\pi r_{H}^{2}-2\pi Q^{2}e^{-2\psi_{0}}.\label{eq:entropy}\end{equation}
All these parameters are related by means of the first law of thermodynamics $dM=TdS+\phi dQ$. For a given fundamental
equation $M=M(S,Q)$ we have the conditions for thermodynamic equilibrium
\begin{equation}
T=\frac{\partial M}{\partial S}\ ,\quad\phi=\frac{\partial M}{\partial Q}.\end{equation}
 Therefore, the phase space ${\cal T}$ for this black hole's geometrothermodynamics
is 5-dimensional with coordinates $Z^{A}=\{M,S,Q,T,\phi\}$ and the
fundamental Gibbs 1-form is given by $\Theta=dM-TdS-\phi dQ$. On
the other hand, the space of thermodynamic equilibrium states ${\cal E}$
is 2-dimensional with coordinates $E^{a}=\{S,Q\}$, and is defined
by means of the mapping \begin{equation}
\varphi:\{S,Q\}\longmapsto\left\{ M(S,Q),S,Q,\frac{\partial M}{\partial S},\frac{\partial M}{\partial Q}\right\} \end{equation}
with $\varphi^{*}(\Theta)=0$. Here, the mass $M$ plays the role
of thermodynamic potential that depends on the extensive variables
$S$ and $Q$, and under this representation, the metric structure
in the phase space ${\cal T}$ can be written from equation (\ref{eq:Gmetric})
as \begin{equation}
G=(dM-TdS-\phi dQ)^{2}+(TS+\phi Q)(dTdS+d\phi dQ).\end{equation}
 However, Legendre transformations allow us to introduce a set of
additional thermodynamic potentials which depend on different combinations
of extensive and intensive variables. In particular, it is possible
to consider the entropy representation, $S=S\left(M,Q\right)$ , where
the Gibbs 1-form of the phase space can be chosen as \begin{equation}
\Theta_{S}=dS-\frac{1}{T}dM+\frac{\phi}{T}dQ.\end{equation}
Then, the space of equilibrium states is defined by the smooth mapping
\begin{equation}
\varphi_{S}:\{M,Q\}\longmapsto\left\{ M,S(M,Q),Q,T(M,Q),\phi(M,Q)\right\} ,\end{equation}
 with \begin{equation}
\frac{1}{T}=\frac{\partial S}{\partial M}\ ,\quad\frac{\phi}{T}=-\frac{\partial S}{\partial Q},\end{equation}
 and such that $\varphi_{S}^{*}(\Theta_{S})=0$. In this representation,
the second law of thermodynamics corresponds to the concavity condition
of the entropy function. Additional representations can easily be
analyzed within GTD, and the only object that is needed in each case
is the smooth mapping $\varphi$ which guarantees the existence of
a well-defined space of equilibrium states. However, the thermodynamic
properties of black holes must be independent of the representation. 

Using the entropy representation, equation (\ref{eq:entropy}) let
us write the entropy as the potential

\begin{equation}
S\left(M,Q\right)=4\pi M^{2}-2\pi Q^{2}e^{-2\psi_{0}}.\end{equation}

Therefore, the Ruppenier's metric is given by

\begin{equation}
g^{R}=-8\pi dMdM+4\pi e^{-2\psi_{0}}dQdQ\end{equation}

and the invariant metric under Legendre transformation induced in
$\mathcal{E}$ is calculated using (\ref{eq:invariantg}) as

\begin{equation}
g=-dMdM+\frac{e^{-2\psi_{0}}}{2}dQdQ.\end{equation}

Here, the curvature vanishes, meaning that the GMGHS black holes do
not show any statistical thermodynamic interaction and no phase transition
structure. On the other hand, considering the mass representation,
we have the potential 

\begin{equation}
M\left(S,Q\right)=\sqrt{\frac{S}{4\pi}+\frac{Q^{2}e^{-2\psi_{0}}}{2}},\end{equation}

and the Weinhold's metric is 

\begin{equation}
g^{W}=-\frac{1}{64\pi^{2}M^{3}}dSdS+\frac{Se^{-2\psi_{0}}}{8\pi M^{3}}dQdQ-\frac{Qe^{-2\psi_{0}}}{16\pi M^{3}}dQdS.\end{equation}
Using equation (\ref{eq:invariantg2}) we obtain Quevedo's invariant
metric,

\begin{equation}
g=-\frac{1}{64\pi^{2}M^{2}}dSdS+\frac{Se^{-2\psi_{0}}}{8\pi M^{2}}dQdQ-\frac{Qe^{-2\psi_{0}}}{16\pi M^{2}}dQdS,\end{equation}
but this time, the curvature escalar gives

\begin{equation}
R=\frac{8\pi\left(2\pi SQ^{2}e^{-2\psi_{0}}+4\pi^{2}Q^{4}e^{-4\psi_{0}}-S^{2}\right)}{\left(S+\pi Q^{2}e^{-2\psi_{0}}\right)^{2}\left(S+2\pi Q^{2}e^{-2\psi_{0}}\right)}.\end{equation}

There are no curvature singularities, showing that GMGHS metric has
no extremal black hole configurations or phase transitions \cite{quevedo10}.
However, it is not in accordance with the intuitive expectation that
naked singularities show the limit of applicability of black hole
thermodynamics \cite{quevedo08,quevedo09,quevedo10,banerjee} and although
the GMGHS solution becomes a naked singularity when $M^{2}\leq\frac{1}{2}Q^{2}e^{-2\psi_{0}}$
(i.e. $S\leq0$), $R$ has no singular behaviour there. 

When $S=\left(1+\sqrt{5}\right)\pi Q^{2}e^{-2\psi_{0}}$
or $M^{2}=\left(\frac{3+\sqrt{5}}{4}\right)Q^{2}e^{-2\psi_{0}}$ the
scalar curvature vanishes identically, leading to a flat geometry.
At this point the scalar curvature changes its sign, and it is the
only point with a positive entropy where this happens. A similar situation
appears in \cite{quevedo08} for Reissner-Nordstrom solution and
the author argues that in GTD a phase transition can also be described
by a change of sign of the scalar curvature, passing through a state
of flat geometry. However, for the GMGHS black hole there is no indication
of a phase transition at this point.

\section{Conclusion}

The formalism of Quevedo's GTD indicates that phase transitions would
occur at those points where the thermodynamic curvature $R$ is singular.
As in ordinary thermodynamics, near the points of phase transitions,
equilibrium thermodynamics is not valid and therefore, one expects
that singularities represent critical points where the geometric description
of GTD does not hold anymore and must give place to a more general
approach. 

However, the relation between the singularities of the specific heat
and the thermodynamic curvature calculated with the Quevedo's metric
\cite{quevedo10} is not consistent for the GMGHS black hole. Our
results show that the metric structure of the thermodynamical manifold
for the GMGHS solution does not have curvature singularities, which
is not in accordance with the intuitive expectation that naked singularities,
as the one that appears for this metric when $M^{2}\leq\frac{1}{2}Q^{2}e^{-2\psi_{0}}$,
show the limit of applicability of black hole thermodynamics. 

It is clear that the phase manifold contains information about thermodynamic
systems; however, it is neccesary a further exploration of the geometric
properties in order to understand where is encoded the thermodynamic information. A more detailed investigation along these lines will be
reported in the future.

\emph{Acknowledgments.} This work was supported by the National University of Colombia.


\begin{thebibliography}{References}

\bibitem{key-1} S.W. Hawking, Comm. Math. Phys. 43 (1975) 199. 

\bibitem{key-2}S.W. Hawking and D.N. Page, Comm. Math. Phys. 87 (1983)
577. 

\bibitem{key-3}P.C.W. Davies, Proc. R. Soc. Lond. A 353 (1977) 499. 

\bibitem{key-4}J. Louko and S.N. Winters-Hilt, Phys. Rev. D 54 (1996)
2647. 

\bibitem{key-5}A. Chamblin, R. Emparan, C.V. Johnson and R.C. Myers,
Phys. Rev. D 60 (1999) 104026. 

\bibitem{key-6}R.-G. Cai, J. Korean Phys. Soc. 33 (1998) S477; R.-G.
Cai and J.-H. Cho, Phys. Rev. D 60 (1999) 067502. 

\bibitem{rup79}G. Ruppeiner, Phys. Rev. A \textbf{20}, 1608 (1979),
Phys. Rev. D \textbf{75} 024037 (2007), Phys. Rev. D \textbf{78} 024016
(2007) .

\bibitem{wei1}F. Weinhold, J. Chem. Phys. \textbf{63}, 2479, 2484,
2488, 2496 (1975); \textbf{65}, 558 (1976).

\bibitem{quev07}H. Quevedo , J. Math. Phys. \textbf{48}, 013506 (2007).

\bibitem{gmghs}G.W. Gibbons, Nucl. Phys. \textbf{B207}, 337 (1982);
G.W. Gibbons and K. Maeda, ibid. \textbf{B298}, 741 (1988); D. Garfinkle,
G.T. Horowitz, and A. Strominger, Phys. Rev. \textbf{D43}, 3140 (1991);
\textbf{45}, 3888(E) (1992)

\bibitem{shaoWEn}Shao-Wen Wei, Yu-Xiao Liu, Ke Yang, Yuan Zhong.
arXiv:1002.1553 {[}hep-th{]}

\bibitem{quevedo10}H. Quevedo, A. Sanchez, S. Taj and A. Vazquez.
arXiv:1010.5599 {[}gr-qc{]}, arXiv:1011.0122 {[}gr-qc{]}

\bibitem{quevedo08}H. Quevedo. Gen.Rel.Grav. \textbf{40}, 971-984
(2008).

\bibitem{quevedo09}H. Quevedo and A. Sanchez. Phys.Rev.D\textbf{
79}, 024012 (2009).

\bibitem{banerjee}R. Banerjee, S. Kumar Modak, and S. Samanta. arXiv:1005.4832
{[}hep-th{]}

\bibitem{key-7}R.A. Fisher, Phil. Trans. R. Soc. Lond. Ser. A 222
(1922) 309; C.R. Rao, Bull. Calcutta Math. Soc. 37 (1945) 81. 

\end{thebibliography}
\end{document}